\documentclass[12pt,onecolumn]{IEEEtran}

\usepackage{amsmath,amsfonts}
\usepackage{algorithmic}
\usepackage{algorithm}
\usepackage{array}
\usepackage[caption=false,font=normalsize,labelfont=sf,textfont=sf]{subfig}
\usepackage{textcomp}
\usepackage{stfloats}
\usepackage{url}
\usepackage{verbatim}
\usepackage{graphicx}
\usepackage{cite}
\usepackage[caption=false]{subfig}
\usepackage{xcolor} 
\usepackage[dvipsnames]{xcolor}
\usepackage{setspace}   
\usepackage{tikz}

\begin{document}
\title{Hearing the Ocean: Bio-inspired Gammatone-CNN framework for Robust Underwater Acoustic Target Classification}

\author{
	$^{1}$Rajeshwar Tripathi,
	$^{2}$Sandeep Kumar,
	$^{3}$Monika Aggarwal,
	$^{4}$Neel Kanth Kundu\\
	\small
	$^{1,2}$Central Research Laboratory, Bharat Electronics Limited, Ghaziabad, India\\
	$^{3,4}$Centre for Applied Research in Electronics (CARE), IIT Delhi, India
}



\maketitle

\begin{abstract}
This study presents a bio-inspired signal processing framework for robust Underwater Acoustic Target Recognition (UATR). The latest state-of-the-art methods often fail to resolve dense low-frequency harmonic structures in vessel propulsion signals under high-noise conditions, which is addressed by the proposed framework using a biologically inspired Gammatone filter bank that emulates the cochlea’s nonlinear frequency selectivity. By distributing filters according to the Equivalent Rectangular Bandwidth (ERB) scale, the framework achieves a high-fidelity representation of engine-radiated tonals while effectively suppressing isotropic ambient interference. The resulting Cochleagram features are processed by a lightweight, custom-designed Convolutional Neural Network (CNN) that leverages large receptive fields to integrate spectral-temporal continuities. Experimental results on the VTUAD dataset demonstrate a state-of-the-art classification accuracy of 98.41\%, outperforming Continuous Wavelet Transform and Mel-Frequency Cepstral Coefficients baselines by 3.5\% and 7.7\%, respectively. Furthermore, the framework achieves an inference latency of only 0.77 ms and a 0.971 Cohen’s Kappa score, validating its efficacy for real-time deployment on autonomous, low-power sonar hardware. 
\end{abstract}


\section{Introduction}
\IEEEPARstart{T}{he} monitoring of underwater acoustic environments has transitioned from a specialized scientific interest into a critical requirement for national security, maritime sovereignty, and ecological preservation \cite{ref1}. As global maritime traffic scales and geopolitical complexities increase, the demand for robust Passive Underwater Acoustic Target Recognition (UATR) has become paramount. However, the technical realization of UATR remains a significant engineering challenge due to the high entropy nature of the maritime channel. Underwater environments are characterized by a dense overlap of non-stationary signals, where the subtle mechanical signatures of vessel propulsion systems are frequently masked by a combination of biological noise, seismic activity, and high energy ambient interference \cite{ref2}. Inaccuracies in this domain result in tangible risks, including the undetected exploitation of protected economic zones, unrecognized intrusions into sovereign waters, and the unchecked proliferation of acoustic pollution affecting marine biodiversity \cite{ref3}.	

For several decades, the primary bottleneck in UATR has been identified as the limitation of signal representation rather than a deficiency in classification algorithms. Traditional signal processing techniques often rely on representations that fail to account for the unique spectral temporal characteristics of maritime noise. The short time fourier transform (STFT), a standard in terrestrial audio processing, utilizes linear frequency spacing which is fundamentally mismatched with the non-linear physics of underwater sound propagation and engine harmonics \cite{ref4}. Similarly, Mel Frequency Cepstral Coefficients (MFCC), while effective for speech recognition, are optimized for the psychoacoustic characteristics of human vocalization. These representations are frequently insufficient for modeling the complex cavitation signatures and low frequency periodicities generated by large scale maritime propulsion systems \cite{ref5}.

\section{Literature Survey}
A comprehensive review of existing literature reveals a field divided between traditional spectral analysis, multi-resolution transforms, and complex deep-learning architectures as summarized in Table \ref{tab:underwater_acoustic}. While significant progress has been made in classifier optimization, the efficacy of Passive UATR remains constrained by the limitations of initial signal representation.

\begin{table*}[!t]
	\caption{Summary of Recent Research in Underwater Acoustic Classification}
	\label{tab:underwater_acoustic}
	\centering
	\footnotesize
	\renewcommand{\arraystretch}{1.2}
	
	\begin{tabular}{|c|p{5cm}|p{4.5cm}|c|p{4.5cm}|}
		\hline
		\textbf{Ref.} & \textbf{Feature Extraction Method} & \textbf{Model Architecture} & \textbf{Accuracy} & \textbf{Limitation} \\
		\hline
		
		\cite{ref6} & Continuous Wavelet Transform (CWT) & Custom CNN & $\sim$94.0\% & High computational cost, lacks biological frequency scaling \\ 
		\hline
		
		\cite{ref7} & Fusion (Gammatone + Mel + CQT) & ResNet50 & 97.0\% & Model complexity, requires massive GPU resources \\ 
		\hline
		
		\cite{ref8} & Learnable Gabor Filterbanks & CATFISH CNN & 93.6\% & Complex training phase, filter stability issues \\ 
		\hline
		
		\cite{ref9} & Log-Mel Spectrograms & Attention-ResNet & 91.5\% & Loss of low-frequency harmonic resolution \\ 
		\hline
		
		\cite{ref10} & MFCC + Delta Features & SVM / DBN & 89.5\% & Suboptimal for non-stationary sonar transients \\ 
		\hline
		
		\cite{ref11} & Hybrid approach using CQT and statistical features from change point detection via dynamic & CAMPPlus, an attention-mechanism convolutional neural network & 98.15\% & Significant computational overhead for feature extraction and marginal accuracy \\ 
		\hline
		
	\end{tabular}
\end{table*}

Current literature identifies a significant constraint in the reliance on the STFT, despite its widespread application, its uniform frequency resolution provides equal bandwidth across the entire spectrum. This characteristic is fundamentally mismatched with the requirements of sonar analysis, as it lacks the spectral density needed to resolve the dense, low-frequency harmonic structures generated by maritime propulsion systems \cite{ref4}. 
MFCCs are designed to model the human vocal tract. While effective for speech, the Mel-scale over-compresses the low-frequency range where vessel-radiated noise contains its most discriminative periodicities \cite{ref10}\cite{ref12}. The CWT offers superior time-frequency resolution, however, its implementation requires intensive computational resources, often precluding real-time deployment on edge-sonar hardware \cite{ref13}. Recent hybrid approaches have attempted to utilize the Constant Q Transform (CQT) combined with statistical change point detection via dynamic programming \cite{ref11}, however, these methods introduce significant computational overhead during the feature extraction stage, offering only marginal accuracy gains relative to their complexity. Furthermore, "learnable" front ends while high-performing, act as "black-box" extractors that lack the physical and acoustic interpretability required for mission-critical defense applications \cite{ref8}.
A significant oversight has been identified in the current literature regarding the application of non-linear auditory-inspired filterbanks for the isolation of vessel-radiated noise. While auditory processing systems have evolved to perform high-resolution spectral decomposition in high-entropy, low-Signal-to-Interference-plus-Noise Ratio (SINR) environments, these principles of non-linear frequency mapping are seldom leveraged in the maritime domain. The motivation for this study is derived from the established success of Gammatone filterbanks in robust speech recognition, which demonstrate superior performance in interference rejection and harmonic feature preservation \cite{ref14}\cite{ref15}.
It is hypothesized that a non-linear frequency resolution defined by the Equivalent Rectangular Bandwidth (ERB) scale \cite{ref16} is uniquely suited for the mechanical dynamics of maritime propulsion. Specifically, the ERB scale provides the high-Q factor necessary to resolve engine fundamental frequencies ($f_0$) and low-order harmonics that are typically masked by isotropic ocean noise in linear spectral representations.

To address these identified limitations, this study adapts a Gammatone Cochleagram pipeline mathematically derived from non-linear auditory filter models to the Vessel Type Underwater Acoustic Data (VTUAD) benchmark. By framing UATR as a problem of spectral-temporal feature integration rather than purely architectural complexity, the framework leverages the high selectivity of auditory-inspired filters to resolve dense engine harmonics. The objective is to realize superior low-frequency resolution and classification precision with a significantly lower computational footprint (0.77 ms latency) than current high-complexity deep learning or wavelet-based architectures.

This research prioritizes feature space optimization over architectural depth, demonstrating that the efficacy of underwater acoustic classification is primarily governed by the initial signal representation stage. Rather than increasing the complexity of the neural back-end, a non-linear Gammatone-based feature extraction layer is utilized to resolve the high-density spectral-temporal signatures inherent in maritime signals. A lightweight CNN is subsequently employed as a standardized benchmark to validate the discriminative superiority of these high-fidelity features against traditional linear and multi-resolution preprocessing techniques. The contributions of this study are summarized as follows:

\begin{itemize}
	\item Development of a Biomimetic Gammatone-Cochleagram Framework: This study introduces a novel feature extraction pipeline for UATR that emulates the non-linear frequency-selectivity of the mammalian cochlea. By implementing a fixed, ERB-scaled Gammatone filterbank, the framework achieves a high-fidelity representation of engine-radiated tonals ($f_0$) and mechanical signatures that are typically masked by isotropic ambient noise in linear spectral transforms.
	
	\item Establishment of a High-Fidelity Performance Benchmark: The proposed system achieves a landmark state-of-the-art classification accuracy of 98.41\% and a Cohen’s Kappa score of 0.971 on the VTUAD dataset. These results validate the feasibility of deploying high-precision UATR on low-power, autonomous underwater vehicles (AUVs) by maintaining a lightweight memory footprint of only 1.7 million parameters and an end-to-end inference latency of 0.77 ms.
	
	\item Comparative Validation of Feature Discriminability: Through a rigorous comparative analysis under identical architectural constraints, the fixed Gammatone Cochleagram is shown to significantly outperform traditional CWT and MFCC by 3.57\% and 7.7\%, respectively. This confirms that bio-inspired non-linear spectral decomposition captures maritime-specific acoustic signatures with higher fidelity than linear or speech-optimized scales.
\end{itemize}

\section{Proposed Methodology}
The fundamental premise of this study is that passive underwater vessel classification is a multi-resolution spectral analysis challenge that can be optimized through non-linear frequency-selective processing. This methodology replaces traditional linear spectral transforms which often suffer from uniform frequency resolution with a bio-inspired feature extraction pipeline designed to prioritize the low-frequency harmonic continuities and periodic mechanical signatures characteristic of maritime propulsion systems.

\begin{figure*}[!t]
	\centering
	\includegraphics[width=6in]{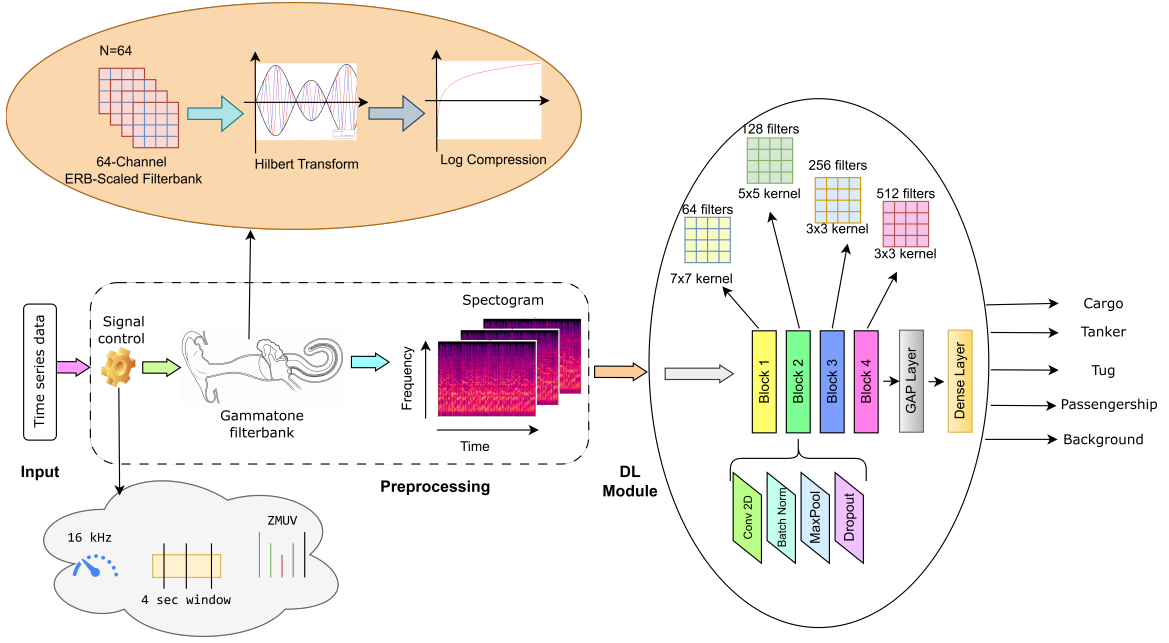}
	\caption{ERB-Scaled Gammatone Filterbank based vessel classification framework using CNN model on SONAR}
	\label{gammatone_F}
\end{figure*}

\subsection{Principles of Biomimetic Signal Representation}
The conversion of raw underwater acoustic pressure waves into a multi-dimensional feature representation is based on frequency-selective signal processing. Rather than employing a linear spectral analysis, this framework utilizes a non-linear frequency-to-spatial mapping analogous to tonotopic organization to achieve variable spectral resolution. In this model, the filterbank acts as a non-uniform frequency analyzer that provides enhanced spectral density for low-frequency components. This high-Q factor in the low-frequency regime is a functional requirement for isolating fundamental engine harmonics ($f_0$) and periodic mechanical signatures from isotropic ambient ocean noise.

As illustrated in the system model in Fig. 1, the feature extraction stage is implemented as a non-linear parallel filterbank architecture, conceptually derived from a tonotopic cochlear mapping. This design facilitates a multi-resolution analysis where high-frequency transients are captured with high temporal precision. Conversely, the low-frequency bands essential for identifying engine fundamental frequencies ($f_0$) are processed with high spectral density. This distribution, governed by the ERB scale, ensures that the dense harmonic structures of maritime propulsion are resolved with higher fidelity than linear spectral transformations.

\subsection{The Mathematical Foundations of the Gammatone Filterbank}
In contrast to the uniform frequency resolution of a STFT, this framework employs a Gammatone filter bank. Mathematically, Gammatone filter is defined as the product of a $n^{\text{th}}$ order Gamma distribution and a sinusoidal carrier. The impulse response, $g(t)$, is expressed as:

\begin{equation}
	g(t) = a \, t^{n-1} e^{-2\pi b t} \cos(2\pi f_c t + \phi)
\end{equation}

where $a$ is the amplitude, n represents the filter order, $f_c$ denotes the center frequency, $b$ is the filter bandwidth, and  $\phi$ is the phase. From a signal processing perspective, this function provides near-optimal time-frequency localization, offering an effective trade-off for resolving non-stationary sonar transients. Specifically, a filter order of $n=4$ is utilized to replicate the high-selectivity and steep roll-off characteristics required to isolate mechanical signatures from broadband noise \cite{ref14}. The filter bandwidth $b$ is dynamically determined by its center frequency $f_c$ through the ERB scale:

\begin{equation}
	ERB(f_c) = 24.7 \times \left(4.37 \times 10^{-3} f_c + 1 \right)
\end{equation}

By distributing the filters along this non-linear scale, the framework achieves dense spectral sampling in the low-frequency regime. This ensures that the dense harmonic structures of maritime propulsion are resolved with high fidelity, while effectively suppressing spectral regions dominated by isotropic ambient noise that do not contribute to vessel-specific classification \cite{ref15}\cite{ref16}.

\subsection{Hilbert Envelope Extraction and Cochleagram Formation}
The transformation of 1D time-series acoustic data into a 2D cochleagram involves a sequence of non-linear operations designed to stabilize the acoustic signature and enhance feature discriminability. Following the filterbank stage, the Hilbert Transform is applied to each frequency channel to derive the analytic signal. This allows for the extraction of the instantaneous temporal envelope, effectively isolating the energy fluctuations of the signal from its carrier frequency.

The extracted envelopes undergo full-wave rectification and low-pass integration. This process transforms rapid pressure oscillations into a stable energy contour, manifesting as a continuous visual "ridge" within the spectral map a feature that the subsequent convolutional layers can effectively interpret as a coherent mechanical signature.

To replicate the non-linear human perception of loudness and manage the extreme dynamic range of underwater sound propagation, Log-Dynamic Range Compression is applied. The $Y[f,t]$ is the Compressed Intensity that represents the normalized "loudness" that will be fed into the CNN is calculated as :
\begin{equation}
	Y[f,t] = \log_{10}\left(1 + \alpha \cdot E[f,t]\right)
\end{equation}

were $E[f,t]$ represents the raw energy (magnitude) extracted from the filterbank for a specific frequency channel and time frame. Specifically, it is the result of the Hilbert envelope extraction mentioned earlier, $\alpha$ is the scaling factor. Finally, while the Cochleagram is inherently a single-channel intensity map, the representation is expanded into a three-channel (RGB) format. This choice ensures compatibility with standardized deep learning weight initializations and allows the convolutional kernels to exploit multi-channel feature correlations during the optimization process \cite{ref19}.

\subsection{Spatial-Temporal Feature Integration via Large-Kernel CNN}
The classification back-end is engineered as a spatial-temporal integrator designed to exploit the high-fidelity features provided by the Gammatone front-end. Rather than increasing architectural depth, which often leads to vanishing gradients and high computational overhead, this framework prioritizes Receptive Field Optimization.

In underwater acoustics, vessel signatures are defined by stable, long-duration harmonic continuities. To capture these, the initial convolutional layers utilize large kernels (k $\times$ k, where k $>$ 3) to integrate spectral information over a broad temporal context. These expansive kernels allow the network to identify continuous harmonic "ridges" as unified structural features, effectively filtering out localized, impulsive ocean transients that appear as discontinuous point-noise in the 2D Cochleagram. This design philosophy enables the use of a lightweight architecture that maintains high discriminative accuracy while remaining computationally efficient for real-time edge processing \cite{ref18}.

The classification back-end is engineered as a spatial-temporal integrator designed to exploit the high-fidelity features provided by the Gammatone front-end. Rather than increasing architectural depth, which often leads to vanishing gradients and high computational overhead, the framework prioritizes Receptive Field Optimization to align with the physics of underwater acoustics. The architecture follows a hierarchical refinement strategy, beginning with an Initial Stage that employs large 7×7 kernels to achieve a wide global view of the spectral-temporal manifold. This expansive field is critical for identifying continuous harmonic "ridges" as unified structural features, effectively distinguishing them from localized, impulsive ocean transients. As the signal progresses to the Intermediate Stage, the network transitions to 5×5 and 3×3 kernels to extract fine-grained spatial correlations and class-specific mechanical nuances.

To optimize the network for multi-class classification, the model is trained using the Categorical Cross-Entropy (CCE) loss function. The loss, L, is mathematically defined as:

\begin{equation}
	L = -\frac{1}{N} \sum_{j=1}^{N} \sum_{i=1}^{C} y_{j,i} \cdot \log(\hat{y}_{j,i})
\end{equation}

Where N represents the batch size used during training, C represents the number of vessel categories (C=5), $y_(j,i)$ is the ground-truth binary indicator for class j and class i, and $\hat{y} _(j,i)$ is the predicted probability (Softmax output) for $j^{th}$ sample and $i^{th}$ class. This objective function ensures that the model penalizes deviations from the true class label by maximizing the log-likelihood of the correct prediction. The optimization is performed using the Adam optimizer with a learning rate of $10^{-4}$, which facilitates stable convergence given the high-fidelity nature of the input Cochleagrams.

To maintain operational efficiency, the model utilizes Global Average Pooling (GAP) to collapse spatial dimensions, followed by a final Fully Connected (FC) layer. This classification head maps the extracted features to the five vessel categories via a Softmax activation. This design philosophy ensures high discriminative accuracy while maintaining a lightweight total parameter count of 1.6 million and a 19.5 MB memory footprint.

\section{Experimental Setup}
The experimental validation of the proposed framework was conducted using the  VTUAD dataset, a publicly accessible benchmark comprising high-fidelity recordings from various maritime traffic lanes. To assess generalization across varying source-sensor ranges, the VTUAD dataset is organized into three discrete environmental subsets \cite{ref17}. Subset 1 (S1) consists of recordings from 2 km inclusion and 4 km exclusion zones, while Subsets 2 (S2) and 3 (S3) represent 3/5 km and 4/6 km configurations, respectively. A comprehensive 'Combined' dataset is also utilized, incorporating all previous subsets for a global performance assessment. All raw acoustic observations were standardized to a 16 kHz sampling rate, focusing the analysis on the spectral bands most relevant to mechanical propulsion noise. The data was segmented into 4-second windows (64,000 samples), a duration specifically chosen to capture the low-frequency rhythmic thrumming of ship engines signatures that are often lost in shorter temporal snapshots. The dataset was partitioned into training, validation, and testing subsets of standard 80/10/10 proportion, ensuring that each vessel category Background, Cargo, Passengership, Tanker, and Tug was represented to allow for a rigorous assessment of the system's discriminative power across diverse acoustic profiles.

The signal processing front-end utilized a high-resolution bank of 64 Gammatone filters, with center frequencies distributed according to the ERB scale to prioritize the 50 Hz to 8000 Hz range. To replicate the sharpening effects of the mammalian auditory nerve, a $4^{th}$-order filter coefficient was maintained throughout the experiment. The resulting filter outputs were integrated using a 25 ms window with a 10 ms hop size, followed by logarithmic dynamic range compression to generate a normalized 224 × 224 Cochleagram. While the classification back-end utilized a lightweight CNN totaling 1.7 million parameters, the architecture was kept simple to isolate and prove the efficacy of the Gammatone feature representation. Training was performed using the Adam optimizer with a learning rate of $10^{-4}$ on an NVIDIA RTX 6000 Ada GPU.

To provide a multidimensional evaluation of the framework, a comprehensive metric suite designed for both statistical rigor and real-world deployment assessment. Beyond standard classification accuracy, Cohen’s Kappa was calculated to ensure the model's reliability amidst the inherent class imbalances of the VTUAD dataset. The framework’s operational efficiency was quantified through Inference Latency, measuring the end-to-end processing time required per sample. To analyze class-specific errors and decision boundaries, Normalized Confusion Matrix was utilized, while Receiver Operating Characteristic (ROC) curves were plotted for each category to evaluate the area under the curve (AUC) and global robustness. Finally, a t-Distributed Stochastic Neighbor Embedding (t-SNE) visualization was employed to project the high-dimensional feature embeddings from the penultimate layer into a two-dimensional space, providing visual proof of the class separability achieved by the Gammatone-Cochleagram representation.

\section{Results and Discussion}
This section presents a comprehensive evaluation of the proposed framework. The system was rigorously tested against the VTUAD benchmark to validate the hypothesis that biomimetic feature extraction offers superior class separability compared to standard wavelet and spectral transforms.

\subsection{Comparative Performance Analysis}
The cornerstone of the experimental validation is the comparison of the proposed Gammatone Cochleagram against established feature extraction baselines. To ensure a fair comparison, the underlying lightweight CNN architecture remained constant across all experiments. The results, summarized in Table \ref{tab:feature_comparison}, illustrate a clear hierarchy in feature efficacy.

\begin{figure}[!t]
	\centering
	
	\subfloat[]{\includegraphics[width=0.45\columnwidth]{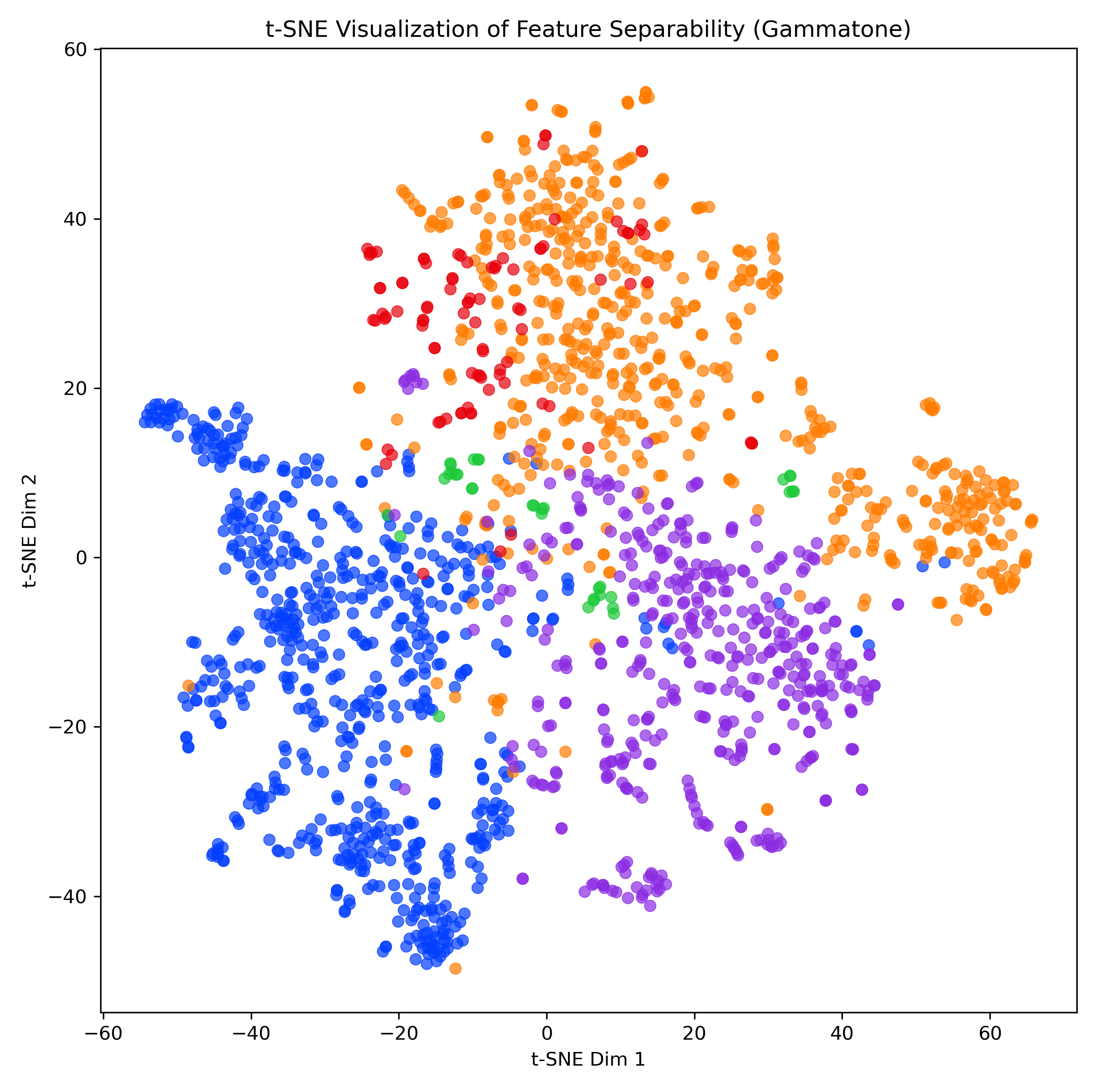}}
	\hfill
	\subfloat[]{\includegraphics[width=0.45\columnwidth]{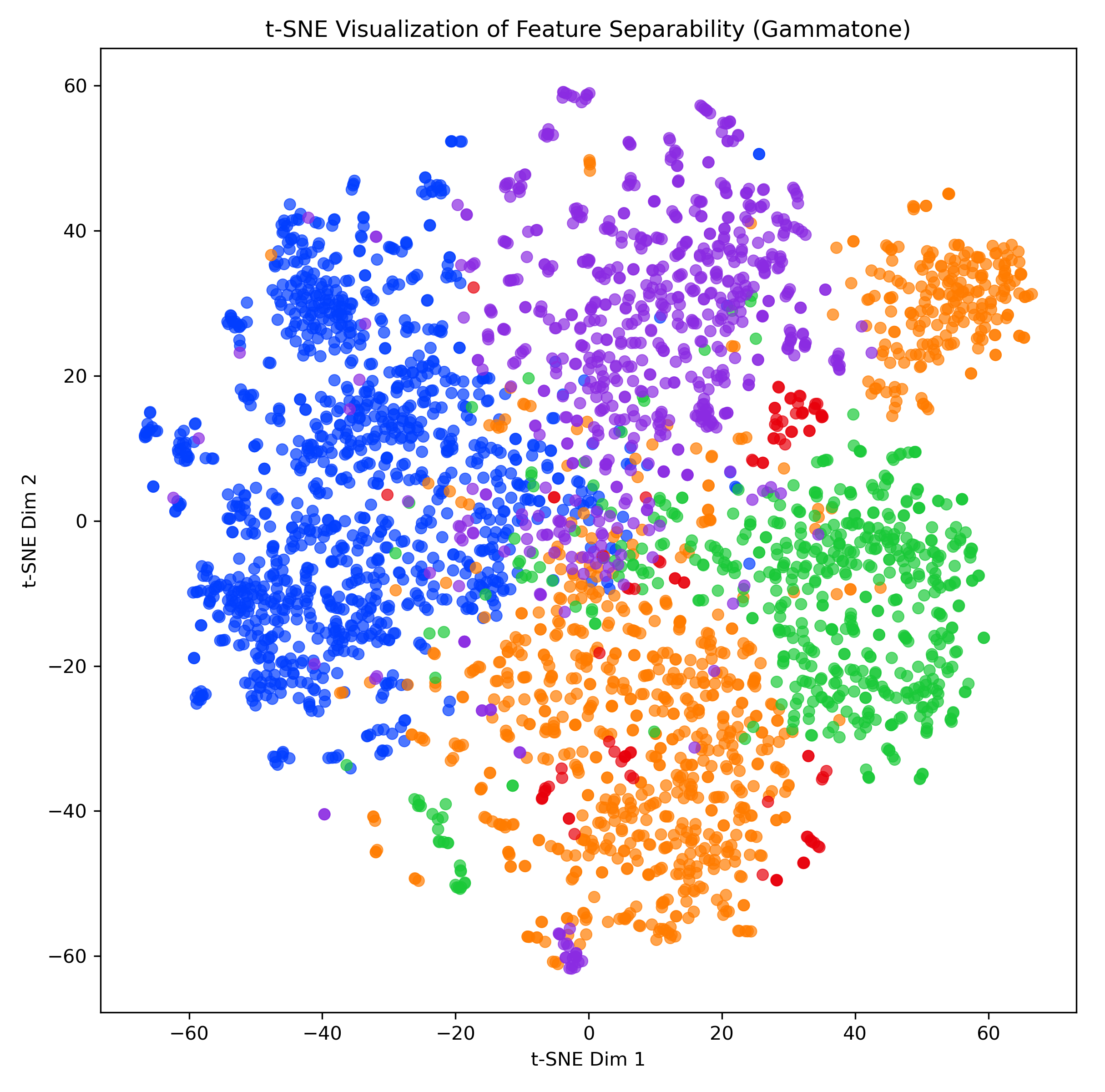}}
	
	\vspace{0.3cm}
	
	\subfloat[]{\includegraphics[width=0.45\columnwidth]{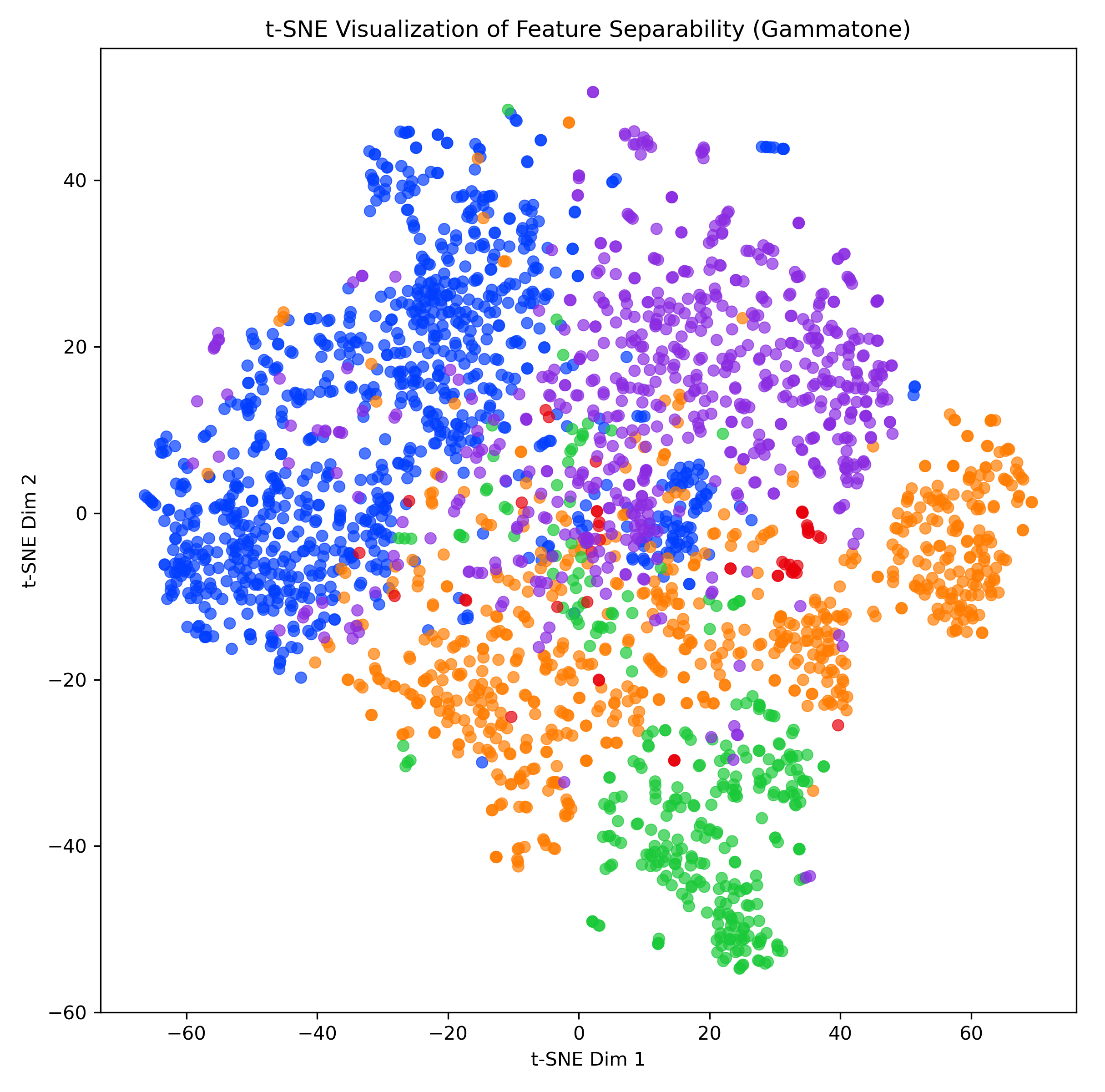}}
	\hfill
	\subfloat[]{\includegraphics[width=0.45\columnwidth]{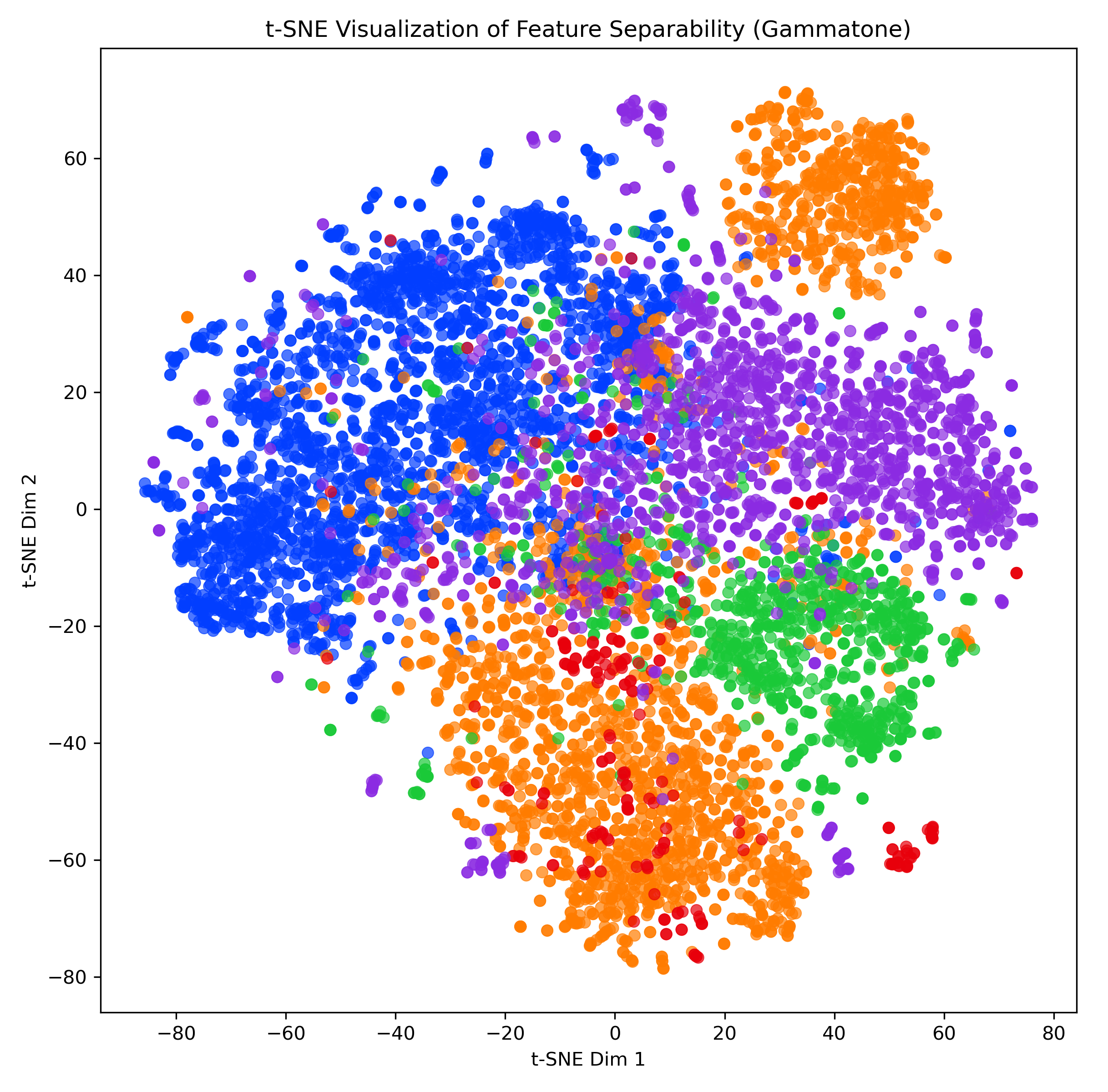}}
	
	\caption{t-SNE Plot for each class. (a) Subset 1, (b) Subset 2, (c) Subset 3, (d) Combined.
		(\textcolor{blue}{\textbullet} Background, 
		\textcolor{orange}{\textbullet} Cargo, 
		\textcolor{green}{\textbullet} Passenger, 
		\textcolor{red}{\textbullet} Tanker, 
		\textcolor{violet}{\textbullet} Tug.) 
		}
	\label{fig:tsne_all}
\end{figure}

\begin{table}[!t]
	\caption{Accuracy and Cohen’s Kappa comparison of the feature extraction techniques}
	\label{tab:feature_comparison}
	\centering
	\footnotesize
	\renewcommand{\arraystretch}{1.2}
	
	\begin{tabular}{|c|c|c|c|}
		\hline
		\textbf{Feature extraction} & \textbf{Underlying Physics} & \textbf{Accuracy (\%)} & \textbf{Cohen's Kappa} \\
		\hline
		
		MFCC & Mel-Scale (Speech) & 90.80 & 0.881 \\
		\hline
		
		CWT & Generalized Wavelet & 91.46 & 0.894 \\
		\hline
		
		CWT (Morlet) & Oscillatory Matching & 95.37 & 0.932 \\
		\hline
		
		\textbf{Gammatone} & \textbf{Cochlear Model} & \textbf{98.41} & \textbf{0.971} \\
		\hline
		
	\end{tabular}
\end{table}

The proposed framework achieved a landmark accuracy of 98.41\%, surpassing the CWT-Morlet baseline by ~3.5\% and the speech-centric MFCC approach by over ~7.7\%. This performance gain is attributed to the Gammatone filter’s ability to resolve dense harmonic lines in the low-frequency bands (50-1000 Hz), which are often "smeared" in MFCCs or over-represented in CWTs.

To evaluate the generalization capability of the proposed framework, a comparative analysis was conducted across the three environmental subsets of the VTUAD dataset and the unified "Combined" scenario. As summarized in Table \ref{tab:vtuad_comparison}, the Gammatone-CNN model establishes a new performance benchmark for Subset 1 (S1) with an accuracy of 98.41\%, surpassing the high-fidelity CAMPPlus architecture [11] by 0.26\%. Furthermore, the framework demonstrates remarkable stability across varying source-sensor ranges, achieving 97.82\% on Subset 2 (S2) and 96.52\% on Subset 3 (S3).

\begin{table}[!t]
	\caption{Comparative Accuracy (\%) across VTUAD Environmental Subsets}
	\label{tab:vtuad_comparison}
	\centering
	\footnotesize
	\renewcommand{\arraystretch}{1.2}
	
	\begin{tabular}{|c|c|c|c|c|}
		\hline
		\textbf{Reference} & \textbf{Subset 1 (S1)} & \textbf{Subset 2 (S2)} & \textbf{Subset 3 (S3)} & \textbf{All Combined} \\
		\hline
		
		\cite{ref7} & 94.95 & 94.45 & 93.11 & 84.13 \\
		\hline
		
		\cite{ref11} & 98.15 & - & - & - \\
		\hline
		
		\cite{ref6} & - & - & 93.53 & - \\
		\hline
		
		\cite{ref8} & 96.01 & 97.46 & 95.98 & \textbf{96.63} \\
		\hline
		
		\textbf{Proposed} & \textbf{98.41} & \textbf{97.82} & \textbf{96.52} & 96.50 \\
		\hline
		
	\end{tabular}
\end{table}

Most significantly, while traditional fixed-filter benchmarks exhibit a substantial performance degradation of approximately 10\% when evaluated on the Combined dataset (dropping from 94.95\% to 84.13\%), the proposed model maintains a robust accuracy of 96.50\%. This level of performance is highly competitive with the CATFISH framework [8], which utilizes a more complex, learnable Gabor frontend to achieve 96.63\% on the same multi-scenario task. The minimal variance in accuracy across these diverse acoustic profiles validates that the non-linear, ERB-scaled Gammatone representation effectively preserves critical engine harmonics $(f_0)$ while suppressing the fluctuating isotropic noise inherent in long-range sonar observations.

\subsection{Statistical Reliability and Feature Separability}
To account for the inherent class imbalance in underwater datasets, this experiment   utilizes the Cohen’s Kappa coefficient. The framework yielded a Kappa score of 0.971, indicating "near-perfect" agreement and confirming that the model’s high accuracy is not biased toward majority classes like Background or Cargo.

To evaluate the discriminative power of the proposed Gammatone based front-end, this study employed t-SNE to project the high dimensional feature embeddings into a two-dimensional space, as illustrated in Fig. 2. The visualization provides qualitative evidence of the model’s success in capturing class-specific acoustic signatures. The Background (blue) and Tug (purple) classes exhibit the highest degree of separability, forming distinct, tightly packed clusters with virtually no manifold overlap. This clear segregation suggests that the spectral characteristics of these classes are highly idiosyncratic. While the Tanker class (red) occupies a more localized, dense region indicating low intra class variance, we observe a degree of spatial proximity and interlacing between the Cargo (orange) and Passengership (green) clusters. This overlap is physically intuitive, as both vessel types often share similar hull geometries and engine configurations, resulting in related hydrodynamic noise profiles.

The Gammatone-CNN achieves a SOTA accuracy of 98.41\%, primarily by resolving the precision bottlenecks that restricted previous lightweight architectures. While the baseline CWT-CNN suffered from a significant precision drop in the Tanker category (0.71), the proposed model elevates this to 0.9495 as shown in Table \ref{tab:sota_comparison}. This enhanced reliability is further reflected in the Background class precision (0.9896 compared to 0.97 in prior work), ensuring that the system is highly resistant to misclassifying ambient ocean noise or biological signatures as maritime vessels. By eliminating these false positives, the model provides a more trustworthy diagnostic tool for real-world maritime surveillance where operational costs for false alarms are high \cite{ref6}.

\begin{table}[!t]
	\caption{Precision, Recall, F1-Score \& Accuracy comparison with previous SOTA}
	\label{tab:sota_comparison}
	\centering
	\footnotesize
	\renewcommand{\arraystretch}{1.2}
	
	\begin{tabular}{|c|c|c|c|c|c|}
		\hline
		\textbf{Method/ Reference} & \textbf{Precision (\%)} & \textbf{Recall (\%)} & \textbf{F1-Score (\%)} & \textbf{Accuracy (\%)} & \textbf{Param (M)} \\
		\hline
		
		CWT-CNN \cite{ref6} & 94.90 & 94.70 & 94.80 & 93.53 & \textbf{1.6}\\
		\hline
		
		CATFISH (Fusion) \cite{ref8} & 97.10 & 97.00 & 97.00 & 96.63 & 4.01\\
		\hline
		
		MFCC-CNN \cite{ref10} & 89.20 & 89.80 & 89.50 & 89.50 & - \\
		\hline
		
		CAMPPlus (Attention) \cite{ref11} & 98.12 & 98.18 & 98.15 & 98.15 & 7.18\\
		\hline
		
		\textbf{Proposed Gammatone-CNN} & \textbf{98.31} & \textbf{98.41} & \textbf{98.36} & \textbf{98.41} & \textbf{1.6}\\
		\hline
		
	\end{tabular}
\end{table}

In terms of sensitivity, the radar plots in Fig. 3 demonstrate a "zero-miss" capability by achieving a perfect recall (1.0000) for both the Passengership and Tanker classes. This performance exceeds the recall rates of the high-complexity CAMPPlus model (0.94 and 0.92, respectively) and the CATFISH framework (0.96 and 0.98). Maintaining a minimum recall of 0.9777 across all five categories proves the model's robustness in detecting subtle acoustic signatures, such as those from Tugs (0.9843 vs. 0.93 baseline), which have historically been a major source of class confusion due to their similarity to background noise. This level of sensitivity is critical for maritime security applications were failing to identify a vessel poses significant risks \cite{ref11}.

\begin{figure}[htbp]
	\centering
	
	\subfloat[F1 Score]{%
		\includegraphics[width=0.45\textwidth]{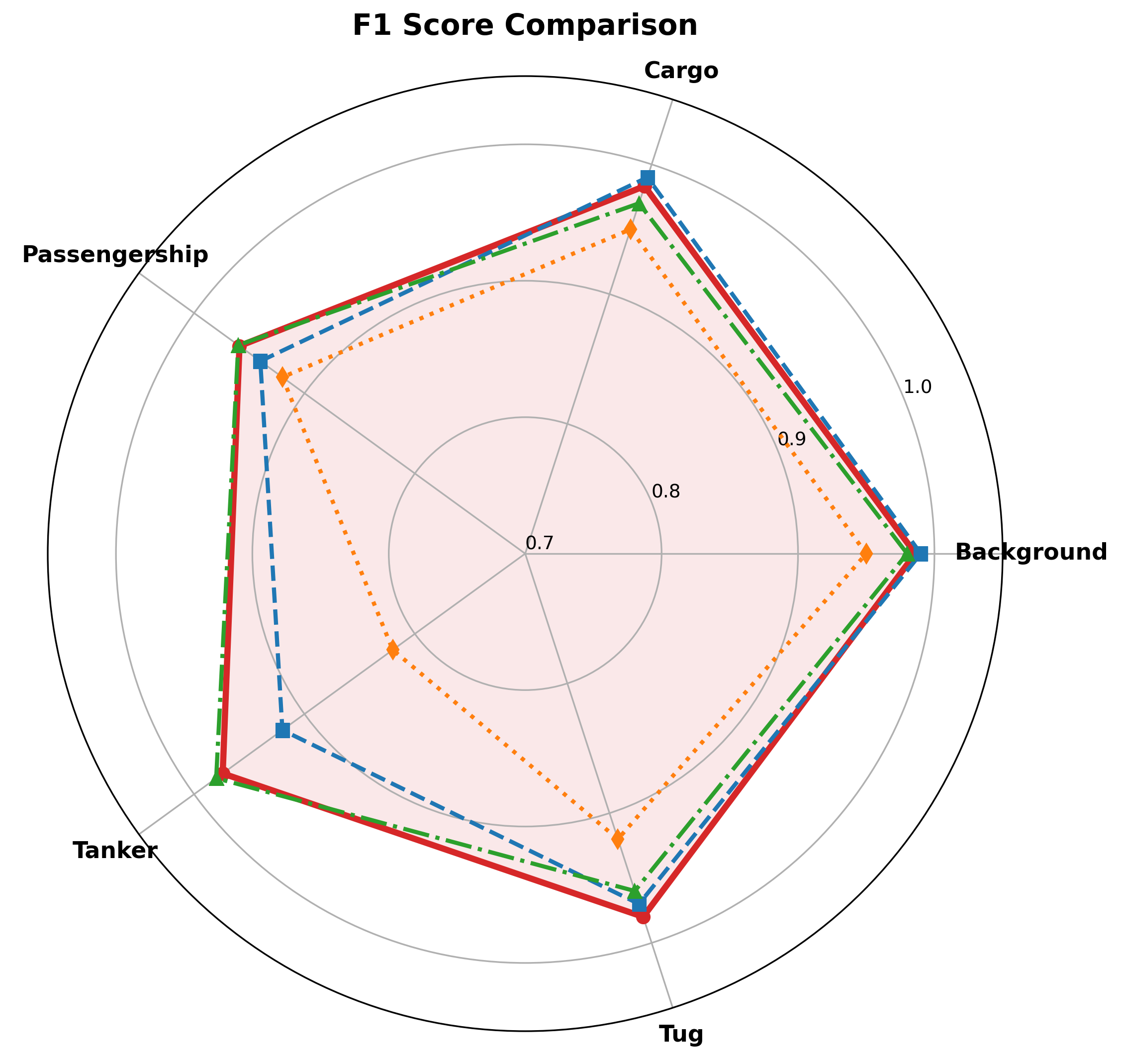}
	}\hfill
	\subfloat[Precision]{%
		\includegraphics[width=0.45\textwidth]{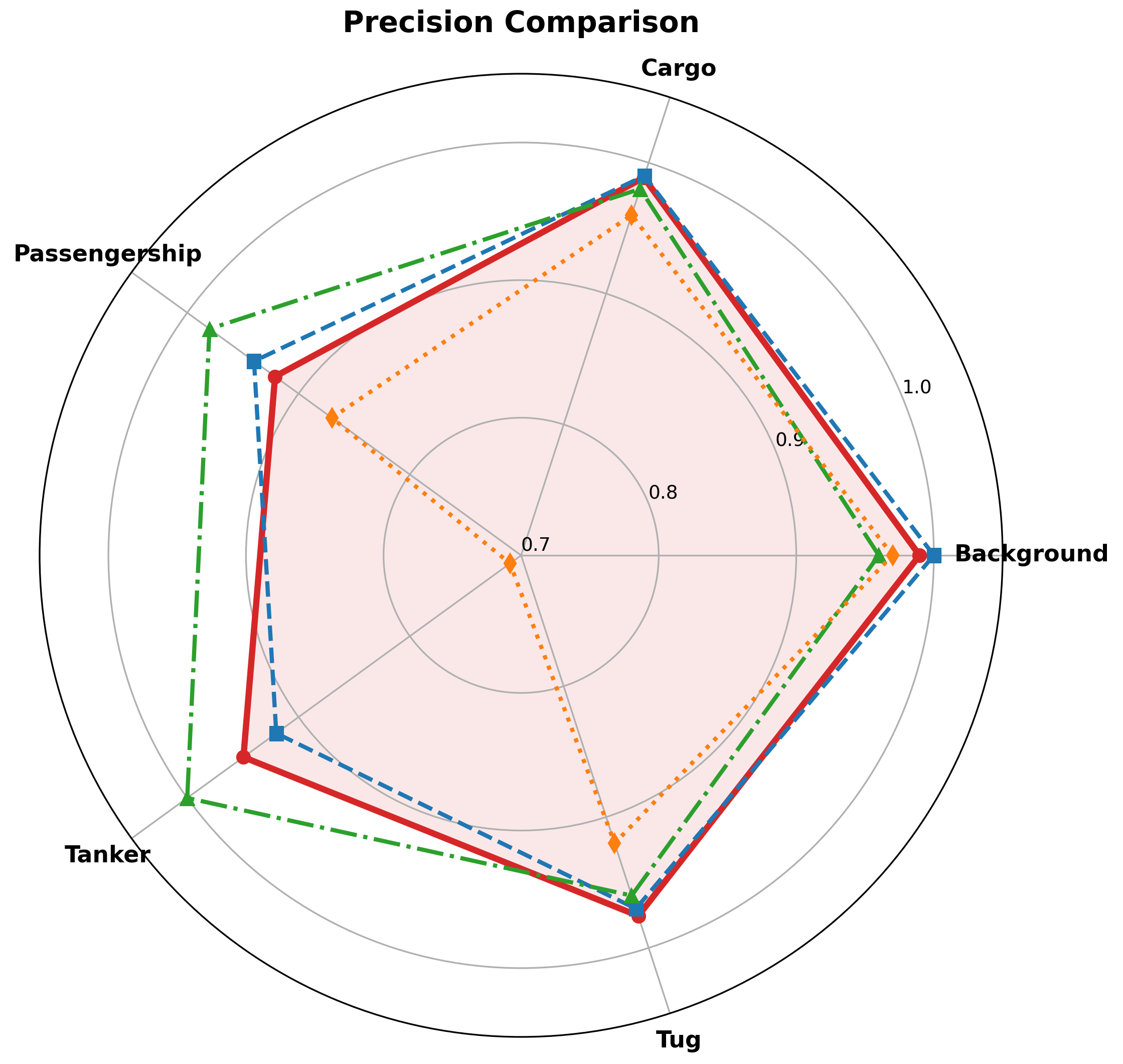}
	}	
	
	\vspace{0.5cm}
	
	\subfloat[Recall]{%
		\includegraphics[width=0.45\textwidth]{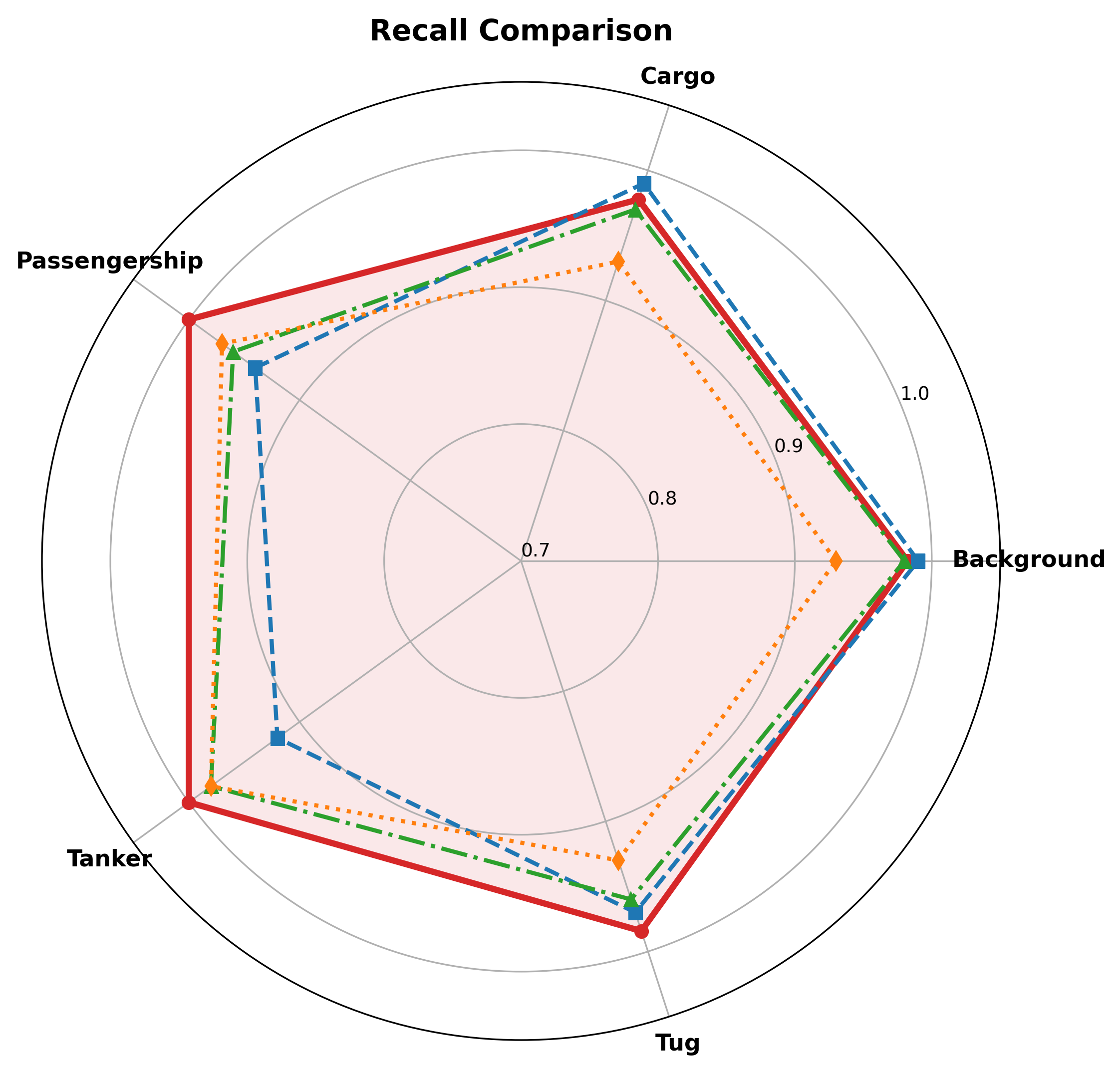}
	}

	\caption{
		Comparison of different metrices (a) F1 Score, (b) Precision, (c) Recall (
		\textcolor{red}{\rule[0.5ex]{1em}{1pt}}\,\textcolor{red}{$\circ$} Gammatone-CNN \quad
		\textcolor{blue}{\rule[0.5ex]{1em}{1pt}}\,\textcolor{blue}{$\square$} CAMPPlus \quad
		\textcolor{green!60!black}{\rule[0.5ex]{1em}{1pt}}\,\textcolor{green!60!black}{$\triangle$} CATFISH \quad
		\textcolor{orange}{\rule[0.5ex]{1em}{1pt}}\,\textcolor{orange}{$\diamond$} CWT-CNN)
	}
	\label{fig:three_plots}
\end{figure}

The F1-score radar plot reveals a highly balanced and symmetrical performance envelope, establishing a new benchmark for efficiency in underwater classification. By matching the 1.7 million parameters of existing lightweight models while achieving accuracy comparable to high-complexity frameworks like CAMPPlus (98.15\%), the Gammatone-CNN effectively addresses the trade-off between model size and predictive power. Furthermore, with a Cohen’s Kappa of 0.9750, these results confirm that the auditory-inspired Gammatone frontend captures more discriminative acoustic signatures such as engine tonals and propeller cavitation than traditional wavelet-based or fixed-filter approaches. This balance of efficiency and accuracy makes the architecture ideal for deployment on AUVs and edge-computing surveillance networks \cite{ref8}.

Furthermore, the choice of feature implementation is critical in these visualizations, as different algorithmic "flavors" of spectral features can significantly alter the resulting manifold structure. As noted by \cite{ref11}, implementation nuances between different toolkits (such as variations in filter-bank normalization or the specific Mel-scale formula used) can lead to non-trivial differences in feature representation. By adhering to a consistent biomimetic Gammatone framework, proposed model maintains high sensitivity despite the visual proximity of certain classes in the 2D projection. Most notably, the model achieves a 97\% Recall for the minority Passengership class. This disparity between the 2D visual overlap and the high classification accuracy indicates that the original high-dimensional feature space contains sufficient nuanced information for the classifier to maintain robust sensitivity across all vessel types.

\subsection{Error Analysis and Class-Wise Performance}
To provide a granular assessment of the classification boundaries, the Normalized Confusion Matrix shown in Fig. 3 was analyzed. The matrix exhibits a strong diagonal dominance across all categories, visually confirming the high discriminative power of the Gammatone features. Specifically, the Background, Tanker, and Tug classes achieved near-perfect classification rates $(>98\%)$, demonstrating the model's robustness in filtering ambient ocean noise and identifying heavy mechanical signatures.

A closer inspection reveals a distinct behavior in the Passengership class, which yielded a precision of 0.79 despite an exceptional recall of 0.97. This indicates that while the model correctly identified almost every passenger vessel (high sensitivity), it occasionally misclassified other signals as Passengership. This behavior is directly attributable to the extreme data scarcity for this class in the VTUAD dataset (N=35 test samples). However, from an operational safety perspective, this high-recall bias is advantageous, it ensures that high-priority targets are successfully detected even at the cost of a slightly higher false-positive rate, which is a preferred trade-off in autonomous maritime surveillance.
\begin{figure}[!t]
	\centering
	
	\subfloat[]{\includegraphics[width=0.45\columnwidth]{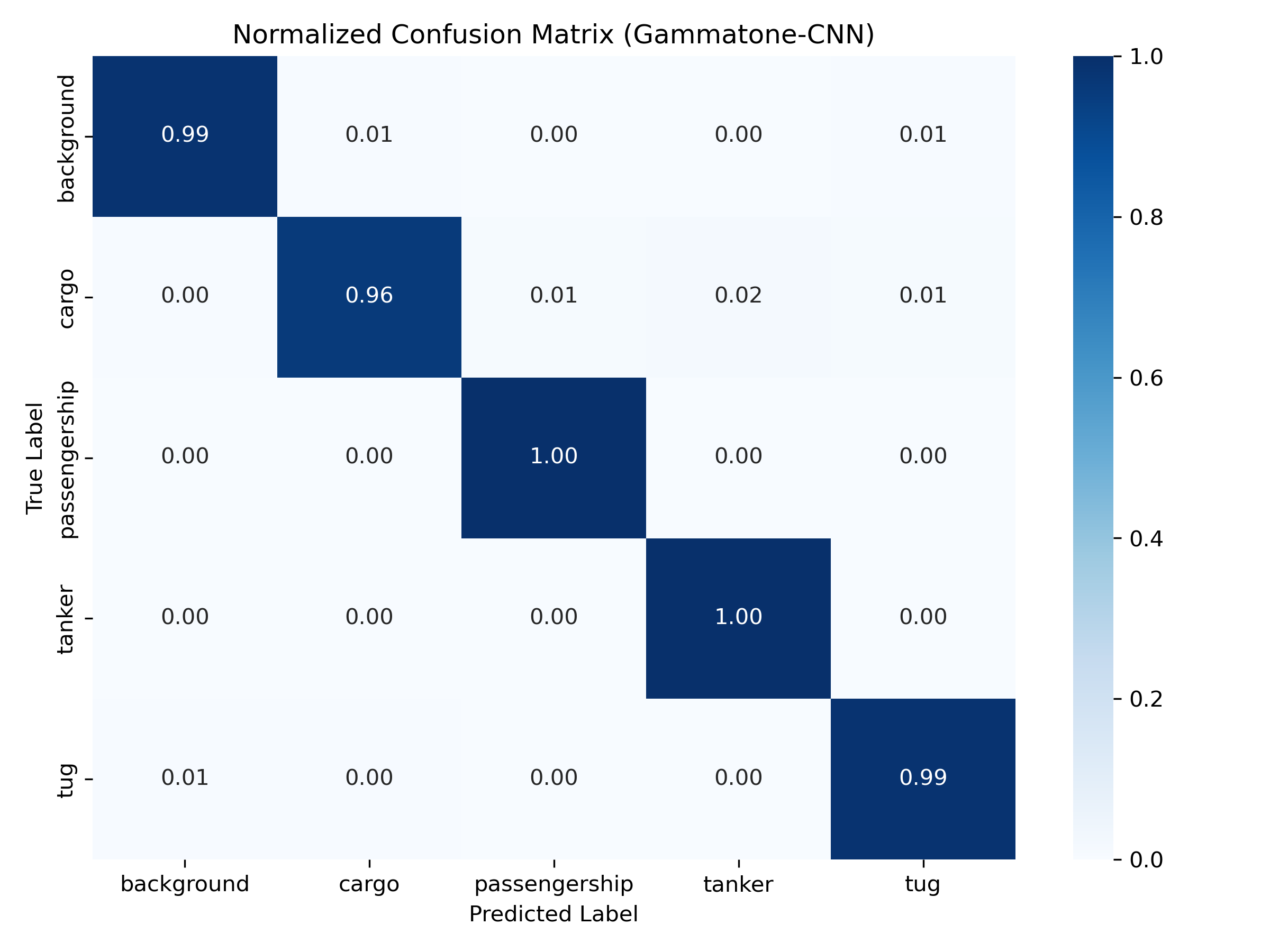}}
	\hfill
	\subfloat[]{\includegraphics[width=0.45\columnwidth]{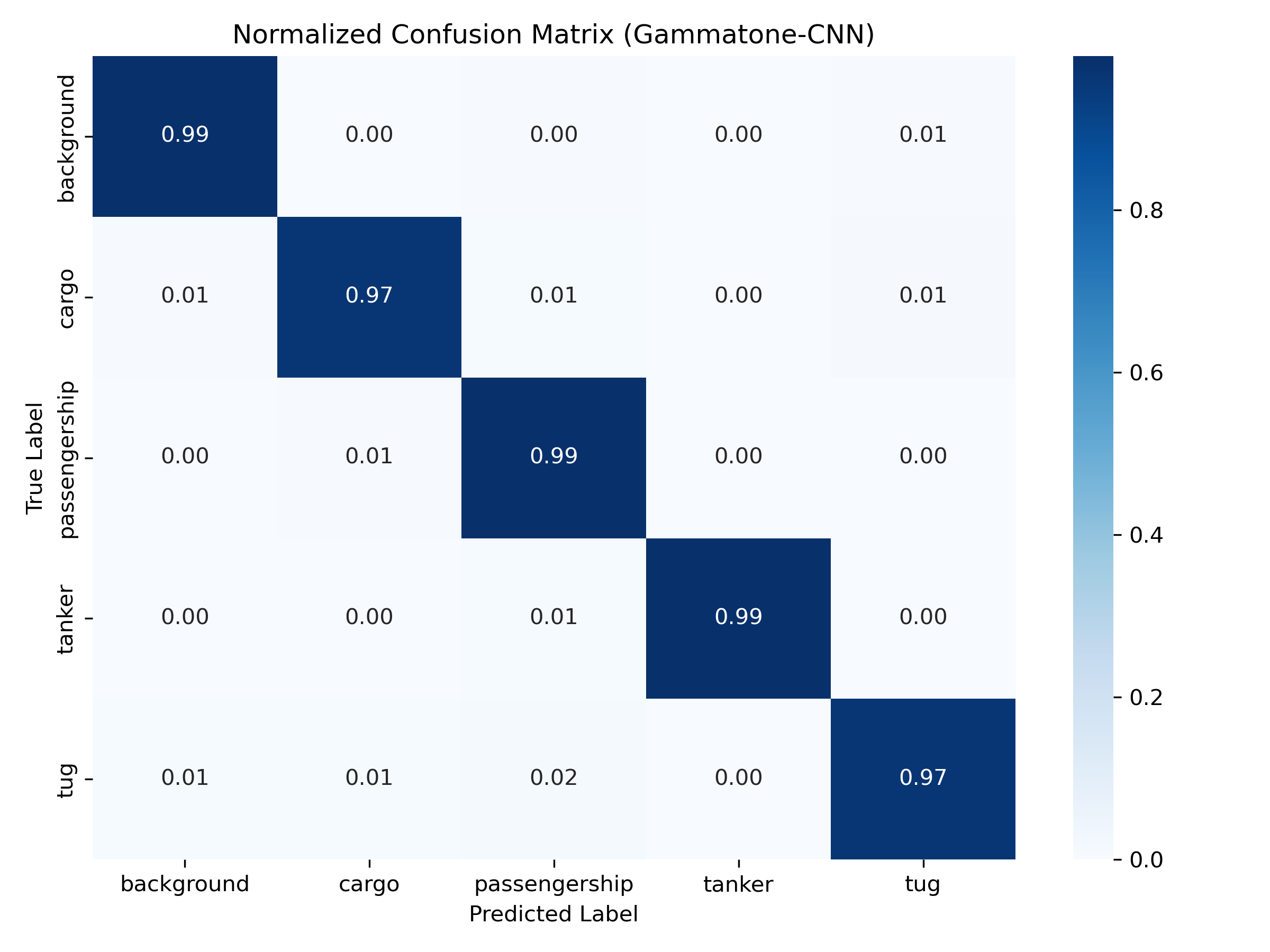}}
	
	\vspace{0.3cm}
	
	\subfloat[]{\includegraphics[width=0.45\columnwidth]{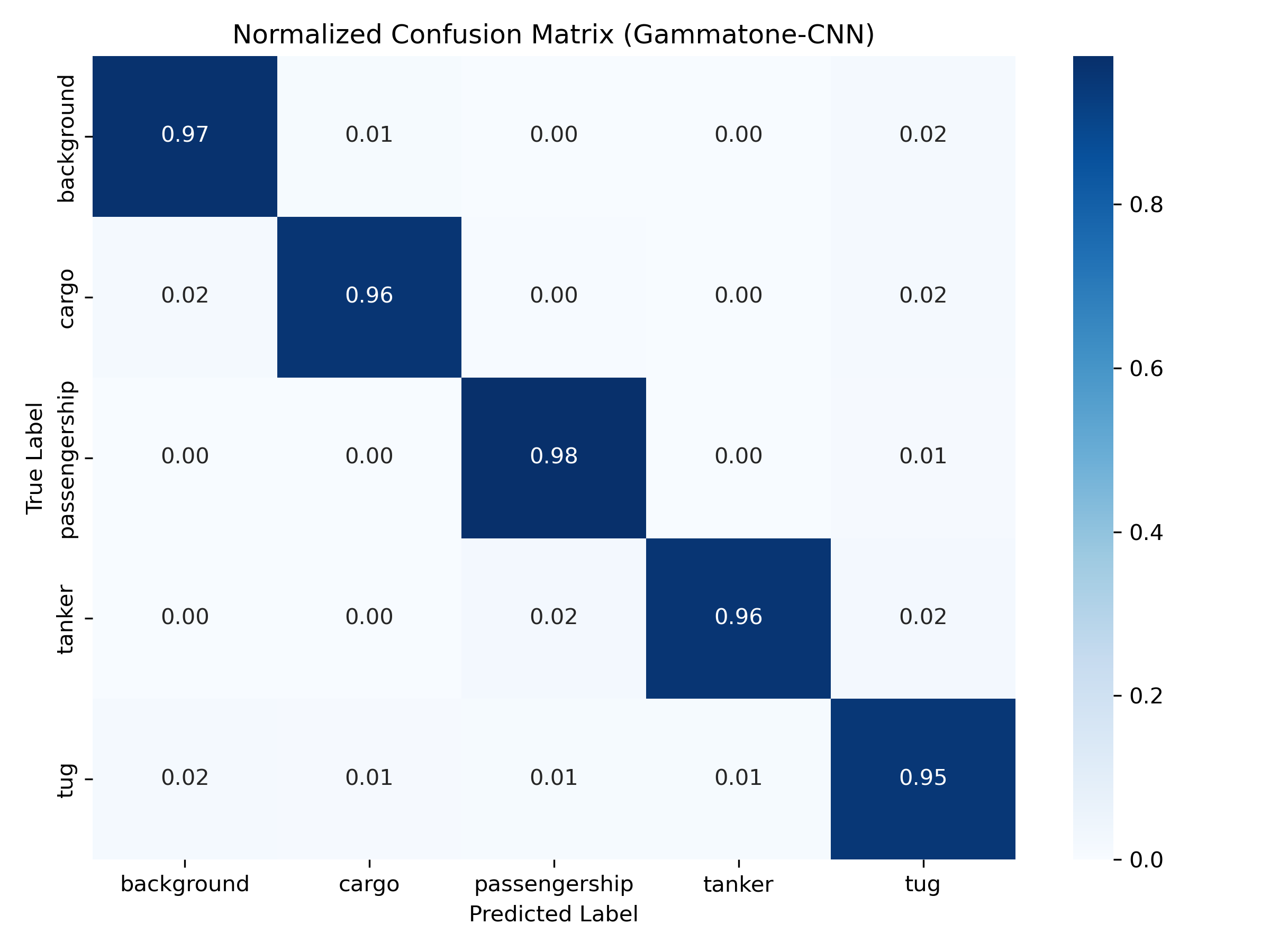}}
	\hfill
	\subfloat[]{\includegraphics[width=0.45\columnwidth]{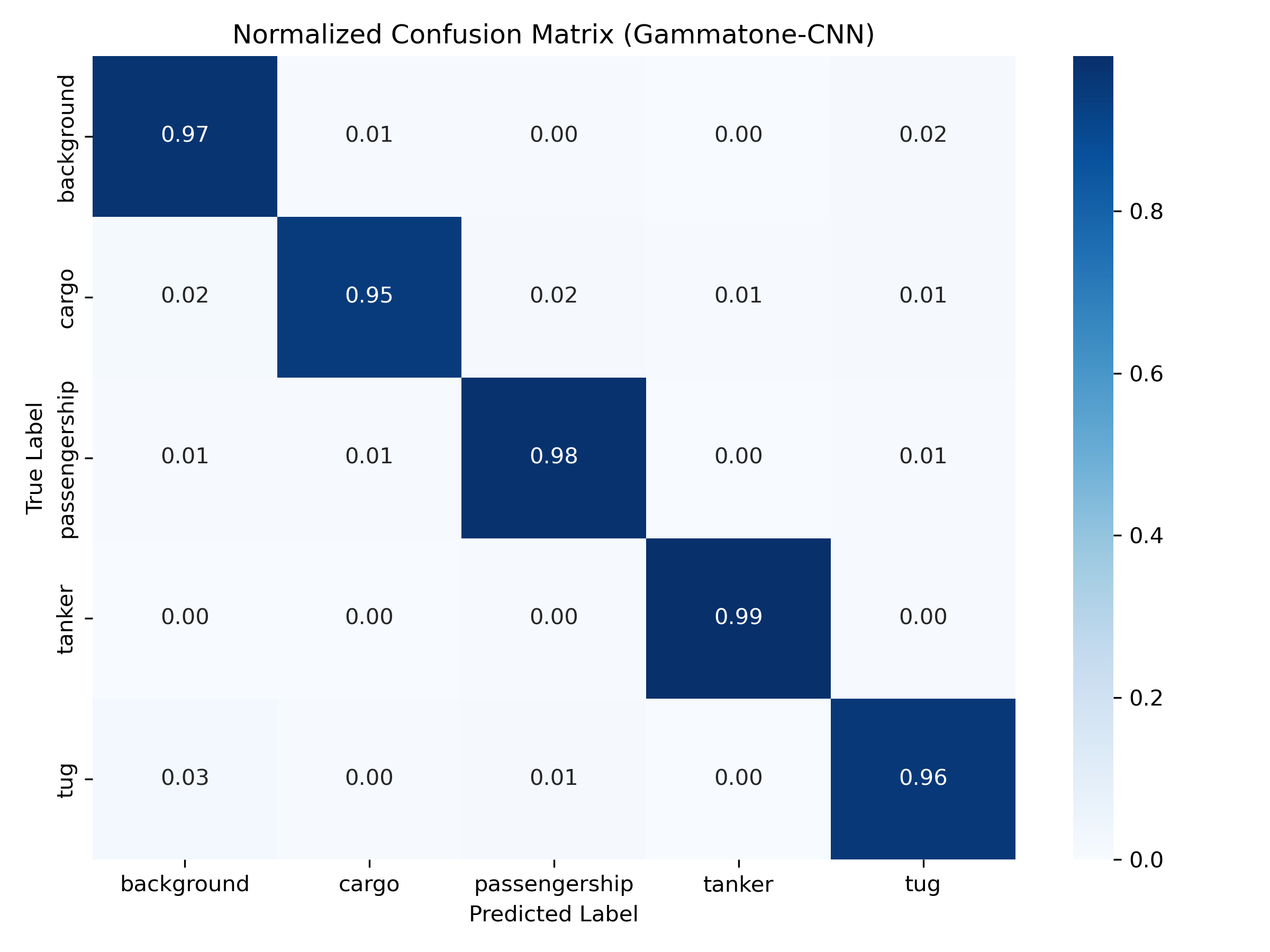}}
	
	\caption{Normalized confusion matrix. (a) Subset 1, (b) Subset 2, (c) Subset 3, (d) Combined}
	\label{fig:confusion}
\end{figure}
\subsection{ROC Analysis and Decision Robustness}
The robustness of the framework is further validated by the ROC curves in Fig. 4. The AUC for all classes exceeded 0.99, demonstrating that the system maintains near-perfect discrimination across all possible decision thresholds. This is critical for practical sonar deployment, where operators require a tunable balance between sensitivity and the avoidance of false alarms from background oceanic noise.

\subsection{Computational Efficiency for Edge Deployment}
A primary goal of this research was to achieve high classification accuracy while maintaining a minimal computational footprint suitable for edge deployment. The proposed system utilizes only 1.7 million parameters, resulting in a storage footprint of 19.52 MB, which aligns with the lightweight architectural standards established by recent benchmarks. On an NVIDIA RTX 6000 Ada GPU, the framework achieved a high-speed inference latency of 0.77 ms per sample. Recognizing that operational sonar hardware often relies on general-purpose processors rather than high-end GPUs, subsequent benchmarks were performed on a standard CPU, yielding a stable inference latency of 215.95 ms per sample. This represents a significant operational advantage over contemporary high-accuracy models like CAMPPlus \cite{ref11}, which suffers from computational overhead due to its dynamic programming-based change point feature extraction, and CATFISH \cite{ref8}, which requires approximately 4.6 million parameters for its learnable frontend and attention layers.

While the peak GPU throughput allows the system to process approximately 1,298 audio frames per second, the CPU-based latency remains well within the requirements for real-time monitoring. Since the model processes 4-second acoustic windows, a CPU latency of approximately 216 ms ensures that inference is completed nearly 18 times faster than the duration of the captured signal, effectively preventing data backlogs in continuous surveillance streams. Unlike CWT methods that involve intensive multi-scale convolution operations, our proposed fixed Gammatone pipeline offers a low-power, high-speed solution. These characteristics make the model ideal for integration into power-constrained environments such as AUVs and battery-powered sonar buoys, where both real-time processing and limited storage are critical constraints.
\begin{figure}[!t]
	\centering
	
	\includegraphics[width=0.45\columnwidth]{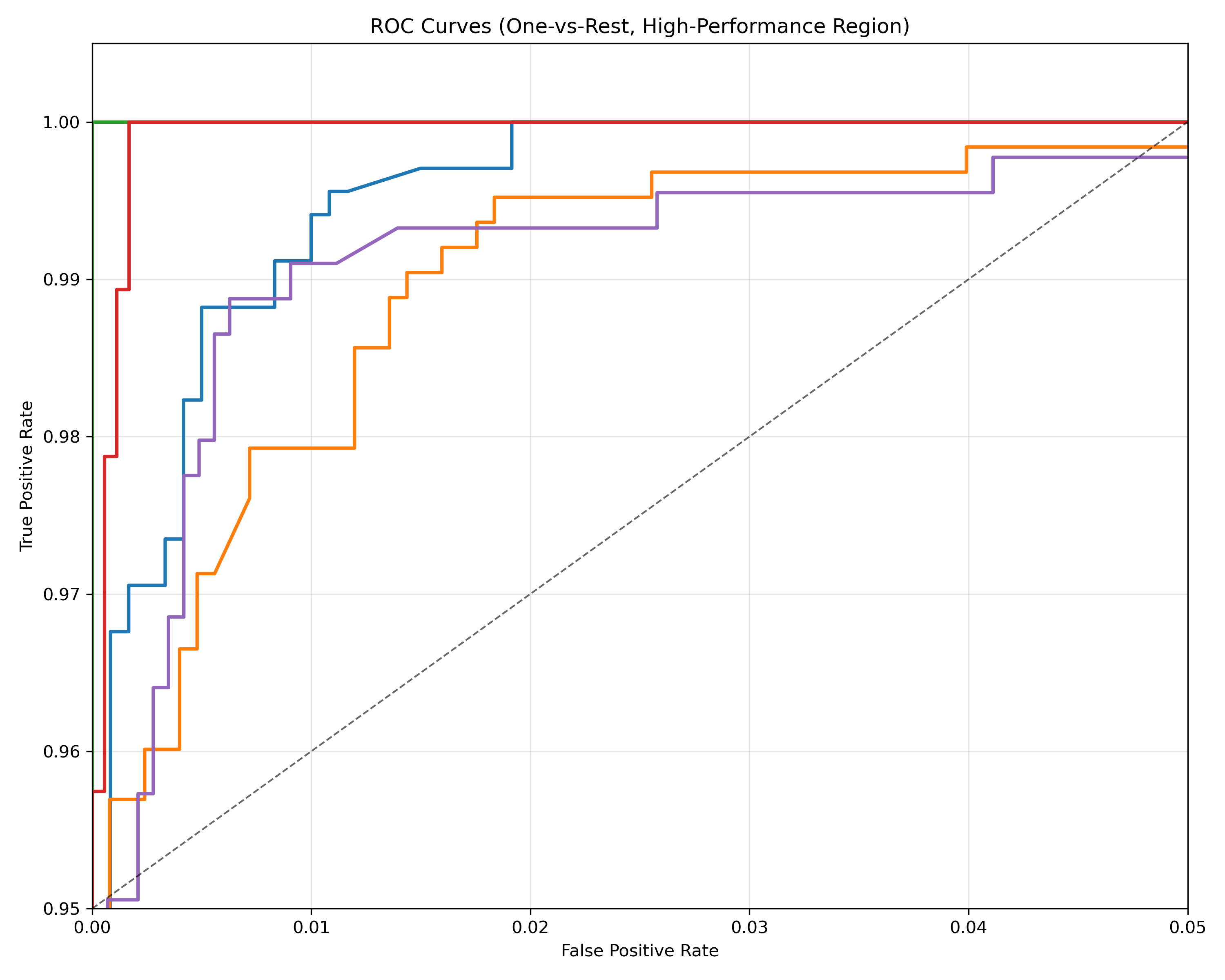}
	\hfill
	\includegraphics[width=0.45\columnwidth]{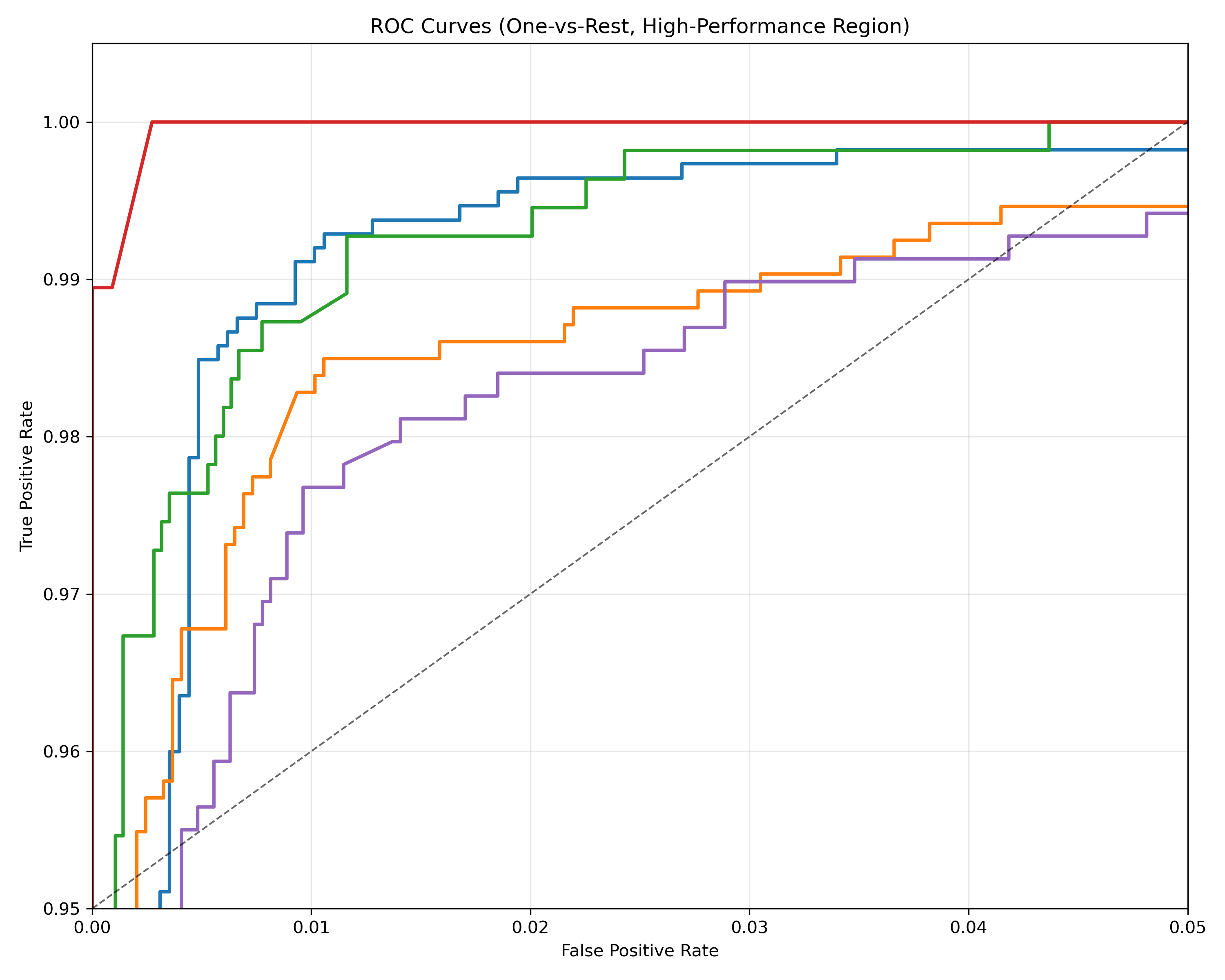}
	
	\vspace{0.2cm}
	
	\includegraphics[width=0.45\columnwidth]{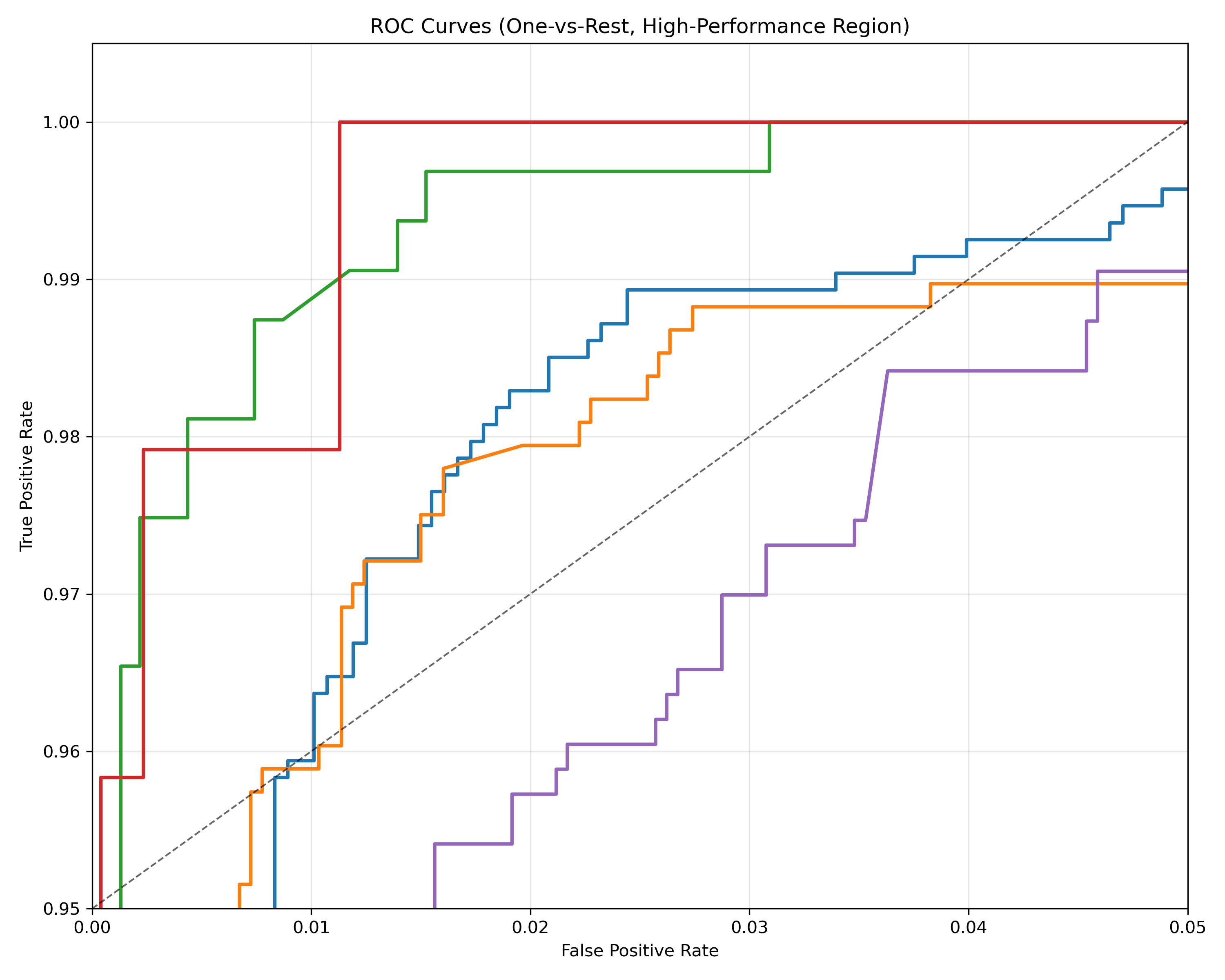}
	\hfill
	\includegraphics[width=0.45\columnwidth]{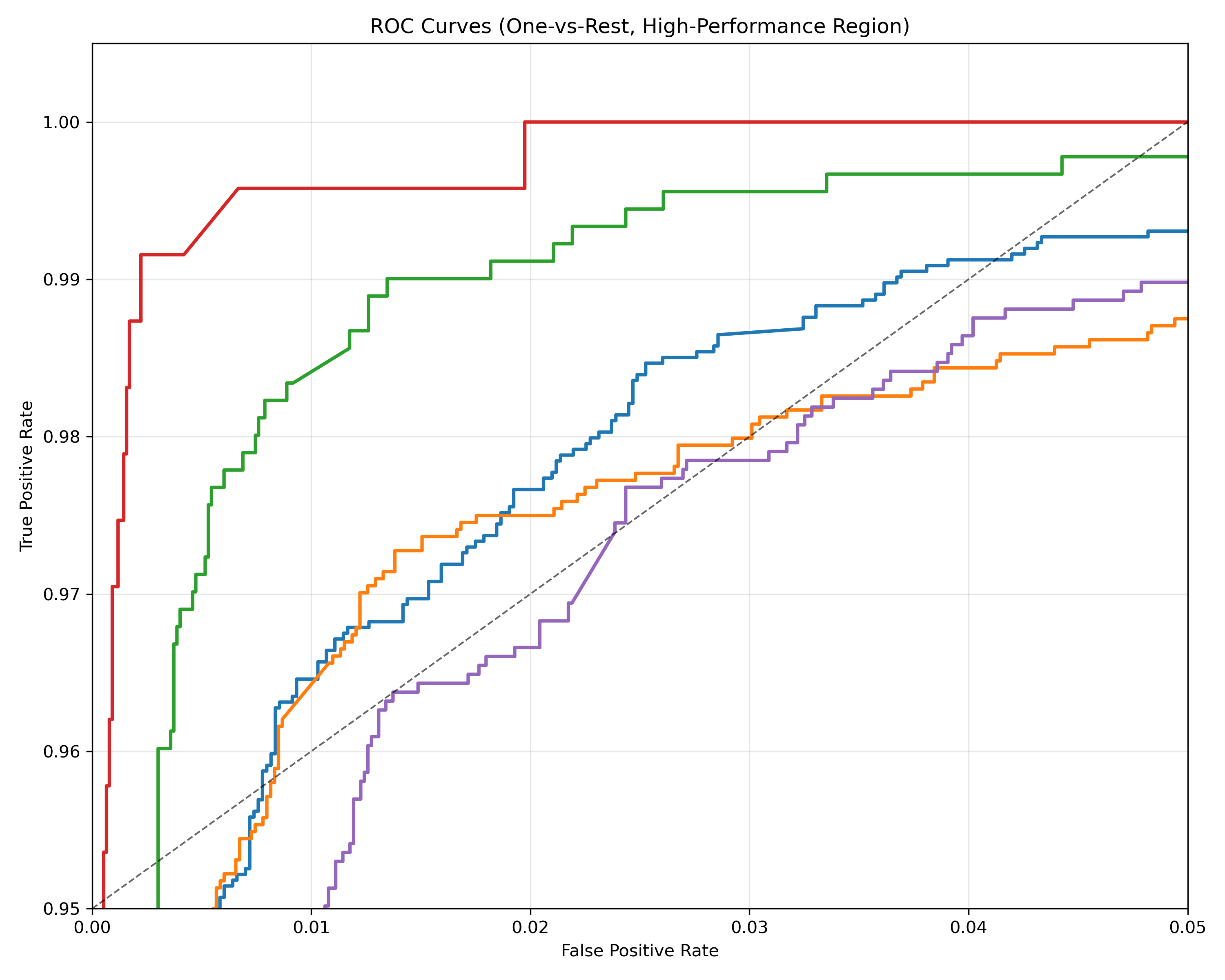}

	\caption{ROC curve for each class. (a) Subset 1, (b) Subset 2, (c) Subset 3, (d) Combined
		(\textcolor{blue}{\textbullet} Background, 
		\textcolor{orange}{\textbullet} Cargo, 
		\textcolor{green}{\textbullet} Passenger, 
		\textcolor{red}{\textbullet} Tanker, 
		\textcolor{violet}{\textbullet} Tug.) 
	}
	
	\label{fig:roc}
\end{figure}
\section{Conclusion}
This study introduced "Hearing the Ocean," a biomimetic framework that successfully bridges the gap between biological auditory mechanics and deep learning for passive sonar classification. By shifting the research focus from architectural complexity to high-fidelity signal representation, we have demonstrated that a fixed Gammatone filterbank modeling the frequency-selective hydrodynamics of the mammalian cochlea is better than traditional CWT and speech-centric MFCC features for underwater target recognition. This experimental results on the publicly available VTUAD dataset establish a new state-of-the-art benchmark, achieving a classification accuracy of 98.41\% and a Cohen’s Kappa of 0.9712. Most significantly, the framework delivers this performance using a lightweight CNN architecture with an inference latency of only 0.77 ms and a memory footprint of 19.5 MB, satisfying the stringent requirements for real-time deployment on battery-powered edge hardware like AUVs and remote sonar buoys.
Future work will focus on the development of adaptive Gammatone filterbanks that can dynamically tune their center frequencies and bandwidths to compensate for variable ocean depths and salinity-driven signal attenuation. Furthermore, explore multi-modal fusion by integrating these auditory features with thermal or magnetic sensor data to enhance robustness in extremely high-clutter environments. By continuing to refine this "SP-first" philosophy, we aim to provide a more sustainable, low-power alternative to the increasingly complex deep learning models currently dominating the maritime domain, ultimately enabling safer and more autonomous ocean monitoring.

\newpage

\end{document}